# Deep Learning Neural Network for Lung Cancer Classification: Enhanced Optimization Function

Bhoj Raj Pandit [1], Abeer Alsadoon[1,2,3,4*], P.W.C. Prasad[1], Sarmad Al Aloussi[5], Tarik A. Rashid[6], Omar Hisham Alsadoon[7], Oday D. Jerew[4]

[1] School of Computing Mathematics and Engineering, Charles Sturt University (CSU), Australia
[2]School of Computer Data and Mathematical Sciences, Western Sydney University (WSU), Sydney, Australia
[3]Kent Institute Australia, Sydney, Australia
[4]Asia Pacific International College (APIC), Sydney, Australia
[5]Computer Technology and Information management Department, Massasoit Community college, MA, USA
[6]Computer Science and Engineering Department, University of Kurdistan Hewler, Erbil, KR, Iraq
[7]Department of Islamic Sciences, Al Iraqia University, Baghdad, Iraq

Abeer Alsadoon[1*]
* Corresponding author. Dr. Abeer Alsadoon, [1]School of Computing and Mathematics, Charles Sturt University, Sydney, Australia, Email: alsadoon.abeer@gmail.com , Phone +61 413971627

## Abstract

**Background and Purpose:** Convolutional neural network is widely used for image recognition in the medical area at nowadays. However, overall accuracy in predicting lung tumor is low and the processing time is high as the error occurred while reconstructing the CT image. The aim of this work is to increase the overall prediction accuracy along with reducing processing time by using multispace image in pooling layer of convolution neural network. **Methodology:** The proposed method has the autoencoder system to improve the overall accuracy, and to predict lung cancer by using multispace image in pooling layer of convolution neural network and Adam Algorithm for optimization. First, the CT images were pre-processed by feeding image to the convolution filter and down sampled by using max pooling. Then, features are extracted using the autoencoder model based on convolutional neural network and multispace image reconstruction technique is used to reduce error while reconstructing the image which then results improved accuracy to predict lung nodule. Finally, the reconstructed images are taken as input for SoftMax classifier to classify the CT images. **Results:** The state-of-art and proposed solutions were processed in Python Tensor Flow and It provides significant increase in accuracy in classification of lung cancer to 99.5 from 98.9 and decrease in processing time from 10 frames/second to 12 seconds/second. **Conclusion:** The proposed solution provides high classification accuracy along with less processing time compared to the state of art. For future research, large dataset can be implemented, and low pixel image can be processed to evaluate the classification.

## *Keywords:*

Convolutional Neural Networks, Auto-encoders, Multispace Image Reconstruction, Deep Learning, Lung cancer, prediction

## 1. Introduction

The rate of death of people by cancer-related disease is increasing daily. Among them, lung tumors are the main cause of death over the world [1]. An early diagnosis is needed. Traditionally, the cancer diagnosis mainly relies on the cancer features like texture, color, and shape of the nodule. These features are problem specific as we could not decide the type of cancer or phase of cancer by looking for color, texture, and shape. Due to the fast development in the deep learning artificial intelligence area, it becomes very effective for image recognition based on color, texture, and shape [2].

The convolutional neural network (CNN), which is part of deep learning in the artificial intelligence. It is very effective in automatic feature extracting. Different algorithms are used for classifying, diagnosing, and analyzing the giving image [2]. Extraction and classification of nodules can be performed by using the CNNs layers, which are a pooling layer, a fully connected layer, and a convolution layer [3]**.** Although CNN is very familiar technology, it also has some constraints that need to be considered, such as the filter size used in CNN, the optimization algorithm, and the image reconstruction process. Many algorithms and techniques are used in CNN to learn and extract the feature so that it can result in better accuracy of feature extraction and in classification of nodules. The classification accuracy is 98.9 % [2]**.** Although the classification accuracy is good, the processing time is higher as the CNN pool uses reconstruction independent subspace analysis (RISA), which is prone to a high processing time and less accuracy. The current solution still needs improvement.

The main objective of the purposed solution is to increase accuracy and reduce the processing time of the system by using the Multispace Image (MIR) method in the pooling layer of CNN [4]. The current RISA solution did not consider the reconstruction loss while calculating overall accuracy. The proposed solution uses MIR, which considered the reconstruction loss by subtracting it from accuracy. Thus, it provides improved accuracy and reduced processing time. The remaining sections of the paper are organized as follow: The “Literature Review” section lists the most important published works in interest area. It also describes the method of the state-of-the-art solution and the proposed method. The “Result” section discusses the various testing techniques used in this research. The “Discussion” section discussed the results obtained from this paper and provides a comparison between the current and the proposed method results. The last section provides the conclusion of the research.

## 2. Literature Review

The primary objective of this section is a survey of the published researches in the area of interest.

Mohamed S et al [5] enhanced the prediction accuracy and minimize the overfitting of cancer features. Mohamed S et al. [5] develop the system that will process the different dimensional data using minimum repetition and wolf heuristic features (optimized neural and soft computing technique). This feature allows selection of the optimized features based on feature selection criteria. Discrete AdaBoost optimized ensemble learning generalized neural networks are used to classify lung nodules and normal features. This improves the error rate by 0.0212 and ensures a high prediction rate (99.48%). For future work, further improvisation can be done by capturing and examining the sensor data and the earlier prediction rate can be enhanced using optimized techniques. Wenqing S et al. [3] developed the multichannel ROI-based deep learning scheme and compare with the performance of a traditional computer-aided diagnosis system (CADx System). The Author design three deep structured algorithms for CNN, Deep Belief Network (DBN), and stacked denoising autoencoders (SDAE). Mohamed S et al. [5] implemented these three algorithms by evaluating each by 10-fold cross-validation. By comparing result of these schemes with the traditional CADx system, Mohamed S et al. [5] found slightly improved area under curve (AUC). The AUC from CNN is $0.899 \pm 0 \cdot 018$ whereas AUC obtained from traditional CADx system is $0.848 \pm 0 \cdot 026$. The research provides future features like flexible input dimensionality, a greater number of layers of algorithms used, and a wide range dataset.”

Ahmed H. et al. [6] evaluated the prognostic signatures using 3D CNN for radiotherapy-treated patients and patients treated with surgery. Ahmed H. et al. [6] used the minimum redundancy and maximum relevance (MRMR) method to select the feature and random forest classifier along with nested cross-validation (5000-fold, 5 times) using caret package. By applying a deep learning approach to patients treated with radiotherapy, Ahmed H. et al. [6] found a 2-year longer mortality rate in patients treated with radiotherapy than in patients treated with surgery. However, the black box nature of deep learning networks means we cannot know the which trained feature is most important. Future research can be done by taking input as multi-dimension and the sensitiveness of medical images are left for future work. Nicolas C, et al. [7] developed a deep learning CNN model (inception V3) to accurately classify whole-slide pathology images into adenocarcinoma, squamous cell carcinoma, or normal lung tissue. Ahmed H. et al. [6] have done classification using random forest classifiers and for optimization. Ahmed H. et al. [6] achieved improved sensitivity and specificity and got an average AUC of 0.97, which automatically outperforms human pathologists. For future work, a visualization tool can be used to identify the feature on the deep learning model. Lakshmanaprabu S.K, et al. [8] applied modified gravitational search algorithm to optimize and classify the computed tomography lung cancer images. Optimal deep learning classifier with multi-gravitational search algorithm (MGSA) optimization algorithm is used to classify the CT lung cancer images. It used for optimal deep neural network and linear discriminate analysis to achieve better accuracy and specificity. Lakshmanaprabu S.K et al. [8] achieve accuracy, sensitivity, and specificity at 96.2%, 94.2%, and 94.56% respectively. This research provides an acceptable range of accuracy, sensitivity, and specificity that is quick, simple, and non-invasive, However, there are also some gaps as Lakshmanaprabu S.K, et al. [8] only used a limited range of data and a single classifier. To get the best solution, Lakshmanaprabu S.K et al. [8] need to use a high amount of data and need to implement multi-classifier. Shiwen S et al. [9] improvise the CNN model by implementing a hierarchical semantic convolutional neural network to provide an interpretable model and high accuracy. Lakshmanaprabu S.K et al. [8] have added a low-level task and a high-level task module that can fine tune the accuracy and present data as an interpretable. It provides better accuracy than the traditional CNN model. The accuracy of 85.6 % with a shorter operating time was achieved. It also presents output as an interpretable form that makes it easier for radiologists to analyze the output. For future work, nodule size, margin speculation, lobulation, labeling error, and anatomic location needs to be considered to achieve higher prediction performance.

Jason L. C et al. [10] developed the NoduleX system, which is a systematic approach for lung cancer nodule malignancy classification. Jason L. C et al. [10] also introduced the quantitative image feature (QIF) extraction. The solution provided in this research provides high accuracy in lung cancer prediction. Here, a nodule <3mm was not considered in this research. In future, to consider these nodules (<3mm), multichannel ROI can be used. Honglin Z et al. [4] proposed the novel framework using MIR. Honglin Z et al. [4] trained the VGG-16 to extract the high-level features of the original image. Honglin Z et al. [10] also introduced the long short-term memory (LSTM) or feature selection and refinement. The classification accuracy is 98.94% in the patch-level classification. Wang C et al. [11] developed the patient-specific adaptive convolutional neural network (A-net) to simulate the radiotherapy. Wang C et al. [11] implemented a population-based neural network for comparison purposes and trained and validated the leave-on-out algorithm. The two algorithms are evaluated based on the coefficient of manual and computerized segmentation like precision. DICE coefficient measures the similarity of a set to another. A-net segmented with a precision, DICE, and root mean square surface distance of 0.81 ± 0.10, 0.82 ± 0.10, and 2.4 ± 1.4 mm, which is far better than a population-based algorithm with 0.63 ± 0.21, 0.64 ± 0.19, and 4.1 ± 3.0 mm, respectively. However, a registration error left unfixed is crucial factor for accuracy. Hence, extensive research needs to be done for determining the reasons behind this result.

Sarfaraz H et al. [1] proposed the system to characterize the nodule by introducing supervised and unsupervised learning modules together. Sarfaraz H et al. [1] introduced a graph regularized sparse MTL platform that helps to integrate the complementary features from lung nodule attributes to improve malignancy predictions. Sarfaraz H et al. [1] provided the best comparison of supervised and unsupervised learning with more accurate results and characterized the nodule in an efficient way. Supervised and unsupervised learning in lung nodule characterization will be helpful as it is providing an acceptable range of sensitivity and accuracy. In future, auto-encoders can be implemented in the unsupervised learning process along with generative adversarial network GANs and multiple instance learning (MIL) can be implemented to analyze medical images. Arkadiusz G et al. [12] constructed the pipeline equipped convolutional network with soft voting to recognize growth pattern of pulmonary adenocarcinoma. Arkadiusz G et al. [12] focused on light CNN architecture, which has low hardware requirements to assess the performance. Arkadiusz G et al. [12] introduced the masking algorithm for reducing the overall analysis time. Arkadiusz G et al. [12] have an accuracy of 89.24%. However, window size is not tested to determine if it decreases the recognition accuracy or not. For implementation of the system on a large scale, further research is needed.

Antonio VAA et al. [2] developed the convolutional neural network model with various autoencoders as the building block of the neural network. Also, Antonio VAA et al. [2] introduce the sparse deep autoencoder to extract the feature. Antonio VAA et al. [2] implement the Adam algorithm for optimization and SoftMax as a classifier. As a result, sparse autoencoder with image input achieves the 98.9% classification accuracy. However, Antonio VAA et al. [2] have a limitation in the pooling layer of the neural network. Antonio VAA et al. [2] have RISA for image reconstruction, which is prone to less accuracy and a high processing time. For the future, low pixel image input needs to be processed for practical accuracy and less processing time.

Lang N et al. [13] enhanced the classification accuracy by introducing dynamic contrast enhance (DCE). Manual ROI was developed to calculate heuristic parameters from wash-in, maximum, and wash-out phase in the DCE. A normalized cut algorithm was introduced to generate a tumor mask and to extract the feature. Radiomics analysis was also performed. The result analyzed for radiomics reached the accuracy of 0.71 and classification using CLSTM improved the accuracy to 0.81. However, the performance cannot be fully relied on as Lang N et al. [13] used a small set of data input. In future, a larger dataset be implemented to investigate better performance in diagnosis.

### 2.1 State-of-the-art Solution

This section presents the features of the current system. In the diagram of the state-of-the-art solution (Figure 1),the content enclosed in the blue broken border shows the favorable features of the current system and the content enclosed in the red broken border shows limitations m. Antonio VAA, et al. [2] developed the CNN model with various autoencoders as the building block of the neural network. Also, Antonio VAA, et al. [2] introduce the sparse deep autoencoder to extract the feature. Antonio VAA et al. [2] implement the Adam algorithm for optimization and SoftMax as a classifier. As a result, sparse autoencoder with image input achieves the 98.9% classification accuracy. Thus, the method proposed by Antonio VAA, et al. [2] has been chosen as the state-of-the-art solution. However, Antonio VAA, et al. [2] have a limitation in the pooling layer of neural network. Antonio VAA, et al. [2] have used RISA for image reconstruction, which is prone to less accuracy and a high processing time. For the future, low pixel image input needs to be processed for practical accuracy and less processing time. As shown in Figure 1, the model is broken into three stages, that is, image preprocessing, feature extraction, and a feature classification stage as shown in block diagram of the state-of-the-art solution. The red color in Figure 1 refers to the limitation of the work and blue color refers to the good feature of it. So, RISA pool is the limitation part, while the preprocessing and classification are the good parts in the state of art solution

*Image Pre-Processing Stage:* The dataset with pathological images from The Cancer Genome Atlas (TCGA) are classified into three transcriptome subtypes [2]. The resolution of original images is over 20,000-40,000 pixels. Antonio VAA, et al. [2] clipped the image into slices of 2048*2048 pixels by feeding the image into the convolutional filter so that the noise in the image is removed for further processing. Thereafter, the filtered image is passed through the max pool, which helps to downsize the image and reduce the dimensions of the image. It extracts the main features from the image better than by average pooling.

*Feature Extraction:* In the feature extraction stage, Antonio VAA, et al. [2] clipped the original image to generate 10,000 sample inputs of size 32*32 pixels. Antonio VAA, et al. [2] implemented the autoencoder model to reconstruct the image and extract the feature of the lung nodule. The pre-processed input is passed through autoencoder, which consists of convolution, RISA pool, deconvolution, and reconstruction layers to extract the feature. Antonio VAA, et al. [2] used a 64*64 image as input and passed it through the convolution filter. The convolution window size 3*3, 5*5, and 4*4 are considered. Antonio VAA, et al. [2] found some improvement in RISA pool when the window size is reduced. The image is reconstructed when the window size is reduced. Antonio VAA, et al. [2] used the three methods to extract features: direct classifier, autoencoder and classifier, and RISA and classifier.

*Limitation:* There is limitation in the pooling layer of neural network. The state of art has used RISA for image reconstruction which is prone to less accuracy and high processing time. While reconstructing the image, large window size filter could not reconstruct efficiently using RISA network, however, this will affect the classification accuracy along with it results high processing time. For future work, low pixel image input needs to be processed to get practical accuracy and less processing time.

*Classification:* Autoencoder is trained to extract the feature from input images and passed through a fully connected dense layer. This layer learns the feature extracted from the autoencoder and classifies the features accordingly. Also, it relates to the SoftMax classifier, which can interpret if the image output is accurate. Antonio VAA, et al. [2] used 128*128, 512*512, and 2048*2048 images as an input and 3*3, 4*4, and 5*5 as a window size for autoencoders, and 16*16 for the classifier. Antonio VAA et al. [2] performed classification in 2048*2048 and found better accuracy. This result means the greater the input image size, the better the accuracy. Antonio VAA, et al. [2] uses the Adam algorithm for optimization and sparsity penalty for the optimization function to reduce the effect of overfitting. Table 1 gives the pseudocode of the state-of-the-art solution. The logical flow diagram of the state-of-the-art system is explained in Figure 2. Antonio VAA, et al. [2] presented the 98.9% accuracy and low cost with high processing time. There is a limitation with this method; the RISA network imposes the reconstruction error. As a result, this method produces less accurate classification output with a high processing time. The optimization function is defined by Equation 1 as follows,

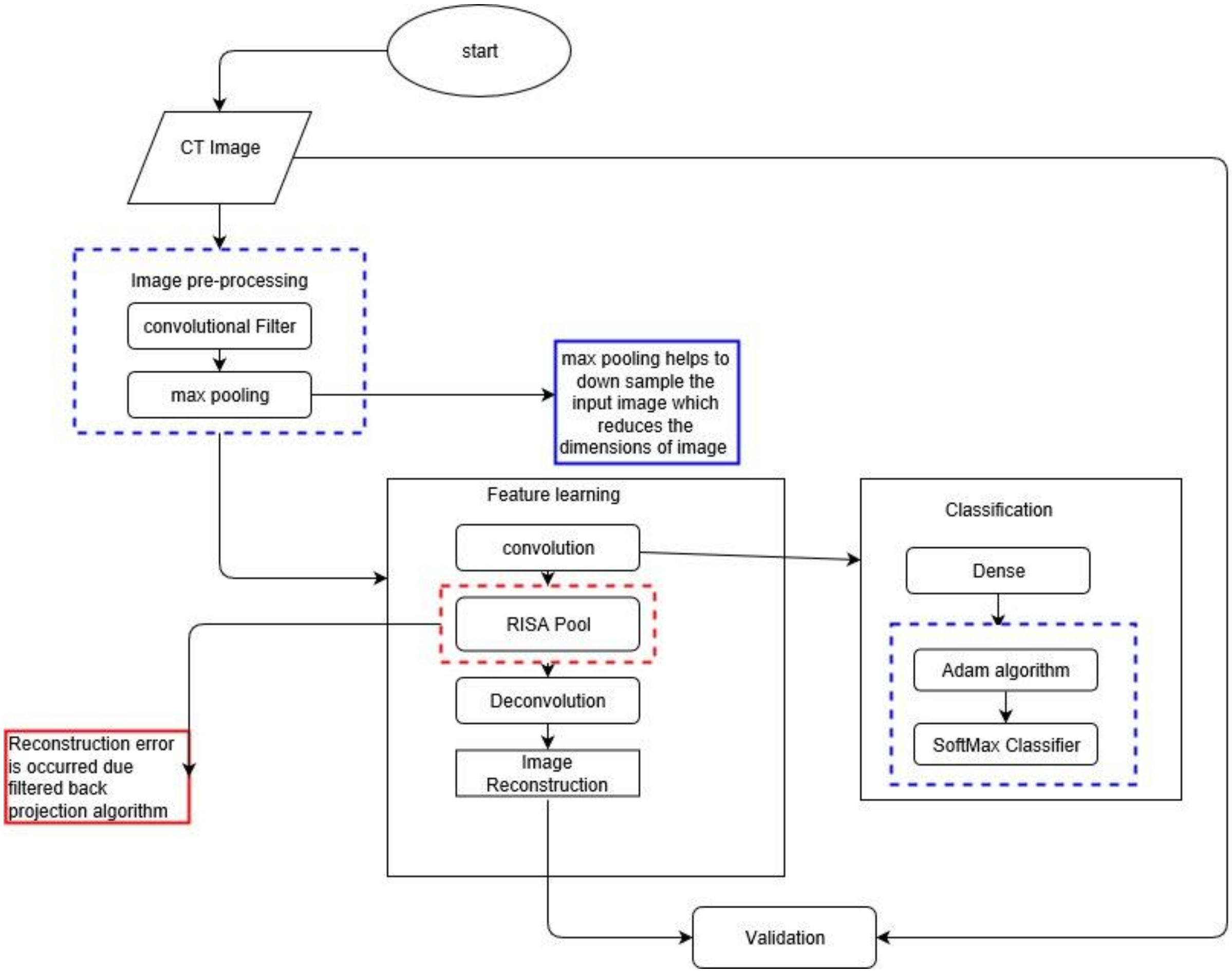

Figure 1: The block diagram of the state-of-the-art system [2]

[The red color refers to the limitation of the work and blue color refers to the good feature of it]

$$L = R + \lambda_s S \quad (1)$$

Where:
L= optimization function
$\lambda_s$ = weight constant

$$R = \sum_{i}^{N} \left(x_i^{out} - x_i^{in}\right)^2 \quad (2)$$

Where:
R = reconstructed function
N= number of nodes in input layer
i = loop initializer
$x_i^{in}$ is the input of layer i
$x_i^{out}$ is the output for given input of layer i

$$S = 1^n \sum_{j=1}^{M} \left(-r_j^{en} \log r_j^{en}\right) \quad (3)$$

Where:
S= sparsity Penalty
M= number of encoding layer
J = encoding layer relating to M
$r_j^{en}$ = output intensity of filter

Table 1: Adam Optimization Algorithm Antonio VAA, et al. [2]

Algorithm: Adam Optimization Algorithm
Input: Raw extracted data
Output: Optimized extracted data

1. Begin
2. Initialize the step size (α)
3. Initialize the hyper-parameters ($\phi_1$ $and$ $\phi_2$)
4. Initialize vector parameter $\theta_0$
5. Initialize first-moment vector ($m_0 = 0$)
6. Initialize Second-moment vector ($v_0 = 0$)
7. Initialize timestep (t=0)
8. While $\theta_t$ not converged do, $t = t + 1$
9. Get gradients $g_t = \nabla\theta ft(\theta_t - 1)$
10. Update 1st biased moment $m_t = \phi_1 \cdot m_{t-1} + (1 - \phi_1) \cdot g_t$
11. Update second biased moment $v_t = \phi_2 \cdot v_{t-1} + (1 - \phi_2) \cdot g_t^2$
12. Estimate bias-correlated first moment $\widehat{m}_t = m_t \,/\, (1 - \phi_1^t)$
13. Estimate bias-correlated second moment $\hat{v}_t = v_t \,/\, (1 - \phi_2^t)$
14. Update $\theta_t = \theta_{t-1} - \alpha \cdot \widehat{m}_t \,/\, \left(\sqrt{\hat{v}_t} + \epsilon\right)$
15. End while
16. Return $\theta t$

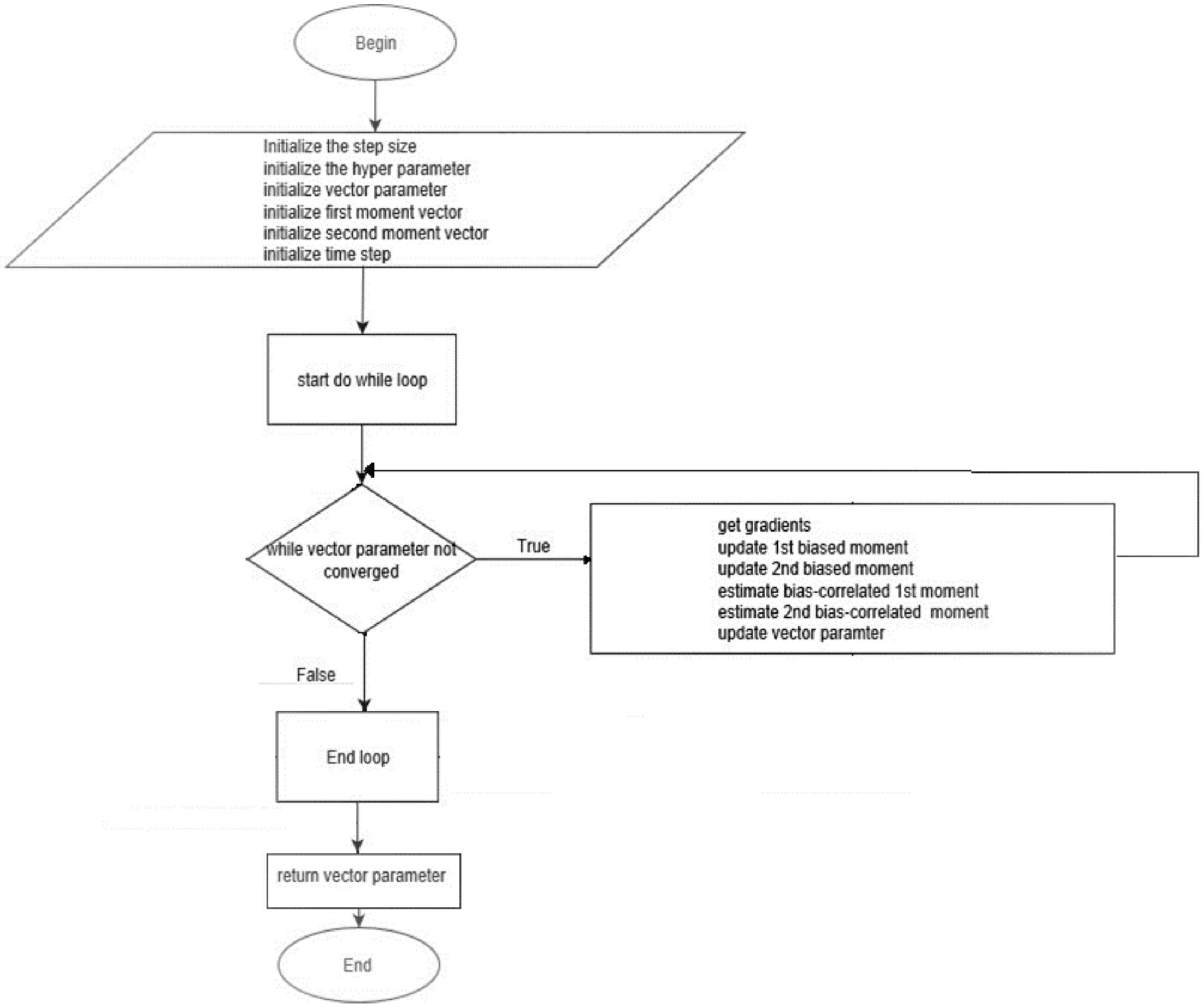

Figure 2: Flowchart of Adam optimization algorithm [2]

# 3. Proposed Solution

We analyzed the advantages and disadvantages of each method. The main issues considered are accuracy, processing time, and quality of image reconstruction.

The best solution, Antonio VAA, et al. [2]**:** use of autoencoder for reduction of dimension and noise. Also, Adam optimization algorithm and SoftMax classifier are favorable features. The Adam optimization algorithm helps to improve the over-fitting effect by reducing the window size. Autoencoders extract the feature with high accuracy. Thus, the Adam optimization algorithm and SoftMax classifier are key to prevent the overfitting effect. However, due to not considering the reconstruction error, the image is not reconstructed well and results in less accuracy in the feature extraction. Also, the use of the RISA tool does not perform well in reconstruction phase. The reconstruction loss is used from the second-best solution [4]. Along with this, we proposed another solution with the MIR tool. This solution considers the reconstruction error while reconstructing the image. The original RGB images were converted into grey level patches and local binary spaces. This information reduces the reconstruction error and the result obtained is more accurate. See Figure 3 for the proposed system. The green borders in Figure 3 indicate the new parts in our proposed system. This means that MIR Pool is the new part that this work has added to the state-of-the-art solution.

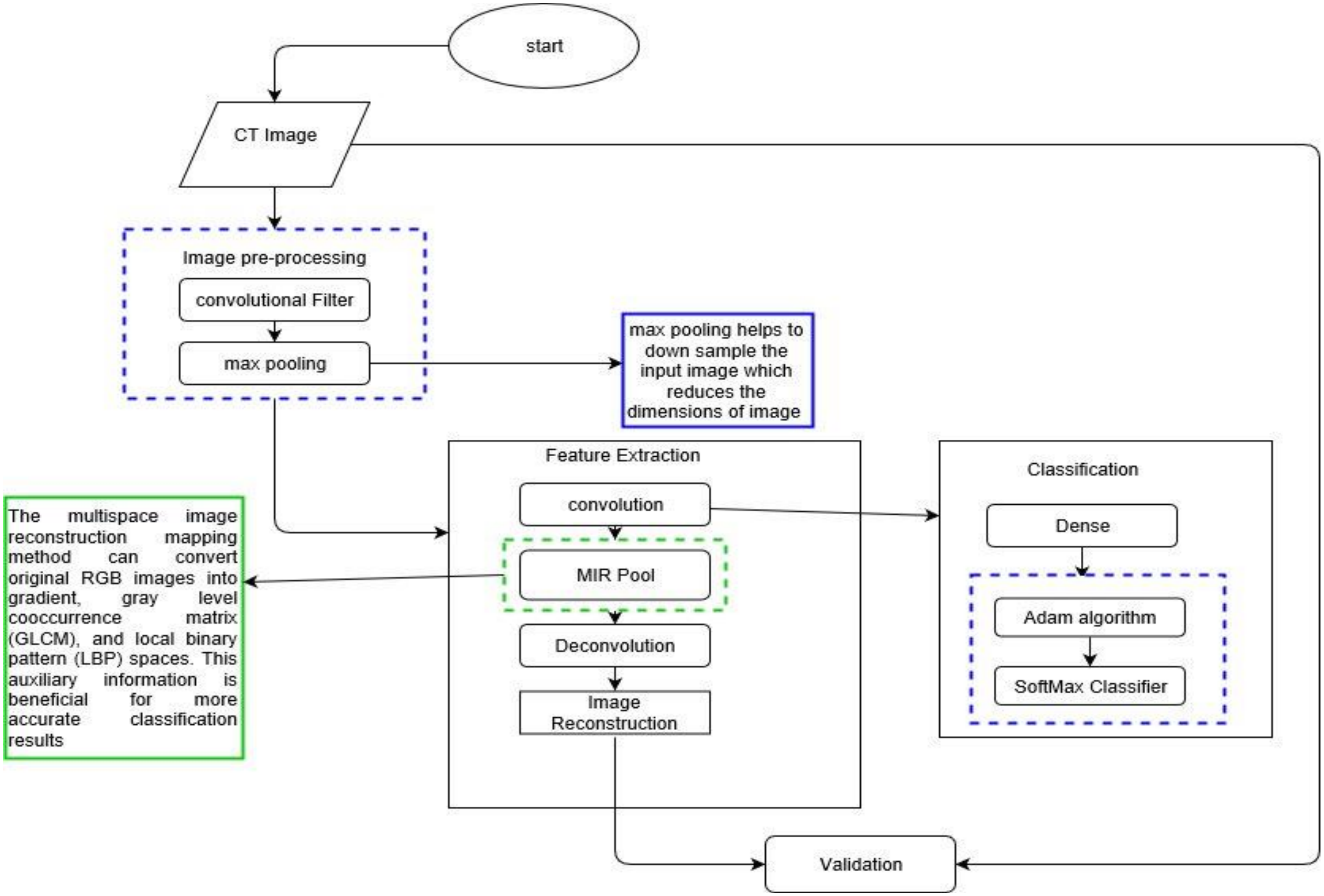

Figure 3: Block diagram of proposed system for decreasing the prediction error,

[The green borders refer to the new parts in our proposed system.]

The proposed system consists of three stages, as shown in Figure 3: Image Pre-Processing stage, Feature Extraction stage, and Classification stage. The green dotted border shows the new feature in the proposed method

*Image Pre-Processing Stage:* The dataset having pathological images from TCGA are classified into three transcriptome subtypes [2]. The resolution of original images is over 20,000-40,000 pixels. Antonio VAA et al. [2] clipped the image into slices of 2048*2048 by feeding the image into the convolutional filter so that the noise in the image is removed for further processing. Thereafter, the filtered image is passed through the max pool, as shown in Figure 3, which helps to downsize the image and reduce the dimensions of the image. Most importantly, it extracts the main features from the image better than by using average pooling.

*Feature Extraction:* In the feature extraction stage, we have clipped the original image to generate 10,000 sample input of size 32*32 pixels. For this, Antonio VAA, et al. [2] implemented the autoencoder model to reconstruct the image and extract the feature of the lung nodule. The pre-processed input is passed through the autoencoder, which consists of the convolution, MIR pool, deconvolution, and reconstruction layers to extract the feature. We used a 64*64 image as input and passed it through the convolution filter. The convolution window size 3*3, 5*5, and 4*4 is considered. We improved the limitation of the state-of-the-art solution, which is the inability of the RISA tool to reconstruct the image. We addressed the reconstruction error in classification accuracy by replacing RISA with MIR.

*Classification:* An autoencoder is trained to extract the feature from input images. Then, the reconstructed image from autoencoder training is passed through the fully connected dense layer, which learns the extracted features and classifies the features accordingly. It relates to the SoftMax classifier, which can interpret the image output accuracy. Antonio VAA, et al. [2] used 128*128, 512*512, and 2048*2048 images as an input and 3*3, 4*4, and 5*5 as a window size for autoencoders, and 16*16 for the classifier. Antonio VAA et al. [2] perform classifications in 2048*2048 and found better accuracy. This result means the greater the input image size, the better the accuracy. Antonio VAA, et al. [2] uses the Adam algorithm for optimization and sparsity penalty for the optimization function to reduce the effect of overfitting.

The reconstruction loss is considered in the proposed solution as defined by Honglin Z, et al. [4]**.** To modify the reconstruction function R to MR, the reconstruction loss needs to subtract from the original reconstruction function defined by Antonio VAA, et al. [2]**.** The modified reconstruction function will be as follows.

## 3.1 Proposed Equation

The state-of-the-art reconstruction function is defined in equation 2. This equation has not considered the reconstructed loss during the reconstruction process. Equation 2 considered the output of the same step to calculate the optimization function, which can only result in a hypothetical value. However, our proposed model considered the output of the previous step to calculate the optimization function.

This process can achieve more classification accuracy. The modified equation is presented below.

$$R1 = \sum_{i}^{N} \left( x_{i-1}^{out} - x_{i}^{in} \right)^2 \qquad (4)$$

Where:
R1 = Modified reconstructed function

N= number of nodes in input layer
if = input layer relating to N
$x^{in}$ is the input
$x_{i-1}^{0ut}$ is the previous output for given input of layer i-1

The Reconstruction Loss RL is defined by Honglin Z, et al. [4] to fix the reconstruction loss, as shown in Equation 5.

$$RL = \sum_i y_i \log b_i \quad (5)$$

Where:
RL = Reconstruction Loss is
$y_i$ = label of each sample of layer i
$a_i$ = original output probability of layer i
$b_i$ =reconstruction output probability of layer i

To find the final modified reconstruction function, the reconstruction loss (calculated from Equation 5) is subtracted from the optimized function (calculated from Equation 1), as in Equation 6. As a result, the reconstruction error will be eliminated, which reconstructs the input image.

$$MR = \sum_i^N \left(x_{i-1}^{out} - x_i^{in}\right)^2 - RL \quad (6)$$

Where:
MR= modified reconstruction function
N= number of nodes in input layer
i = input layer relating to N
$x_i^{in}$ is the input of layer i
$x_{i-1}^{0ut}$ is the previous output for given input of i-1.
RL= Reconstruction loss

Finally, the Enhanced optimization function in our proposed system modified as shown in equation 1 and equation 6 as follows, this will result in better classification accuracy and processing time.

$$EML = MR + \lambda_s S \quad (7)$$

Where:
EML = Enhanced modified optimization function
MR= Modified Reconstruction Function
$\lambda_s$ = weight constant
S= Sparsity Penalty

## 3.2 Area of Improvement

We proposed two equations in our research, which are Equation 4 and 7. With the help of the proposed equations, the reconstruction error is eliminated. The reconstructed image will be obtained clearly and with no blurriness. Furthermore, we have considered the filter size. We have implemented the different window filter size and worked to increase the processing time. To achieve a faster time, we increased the filter size to process the large amount of input data. Why the MIR Tool? Multispacer image reconstruction generates the new image that contains the three channels: gradient, gray level co-occurrence matrix (GLCM), and local binary pattern (LBP). The proposed model considers the best tool for image reconstruction with the ability to remove the drawbacks of the state-of-the-art model. Because the RISA tool cannot handle the color and intensity variations, we proposed the MIR tool as a solution. The image needs to be reconstructed with no blur to achieve the high accuracy and processing time. Because the RISA tool is hampered by color and intensity variations, we cannot achieve an accurate result in such conditions. Also, an increase in the filter window size results in a decrease in accuracy and processing time. To correct these issues, we needed the MIR tool, which can handle the color and improves intensity variations and accuracy. *(Comparison)* As we can see from the state-of-the-art solution and other researchers [5], no other solutions addressed the issue of reduced classification accuracy while increasing the filter window size. No other tool can handle the color and intensity variations. Our proposed solution has eliminated these issues with enhanced classification accuracy and less reconstruction error.

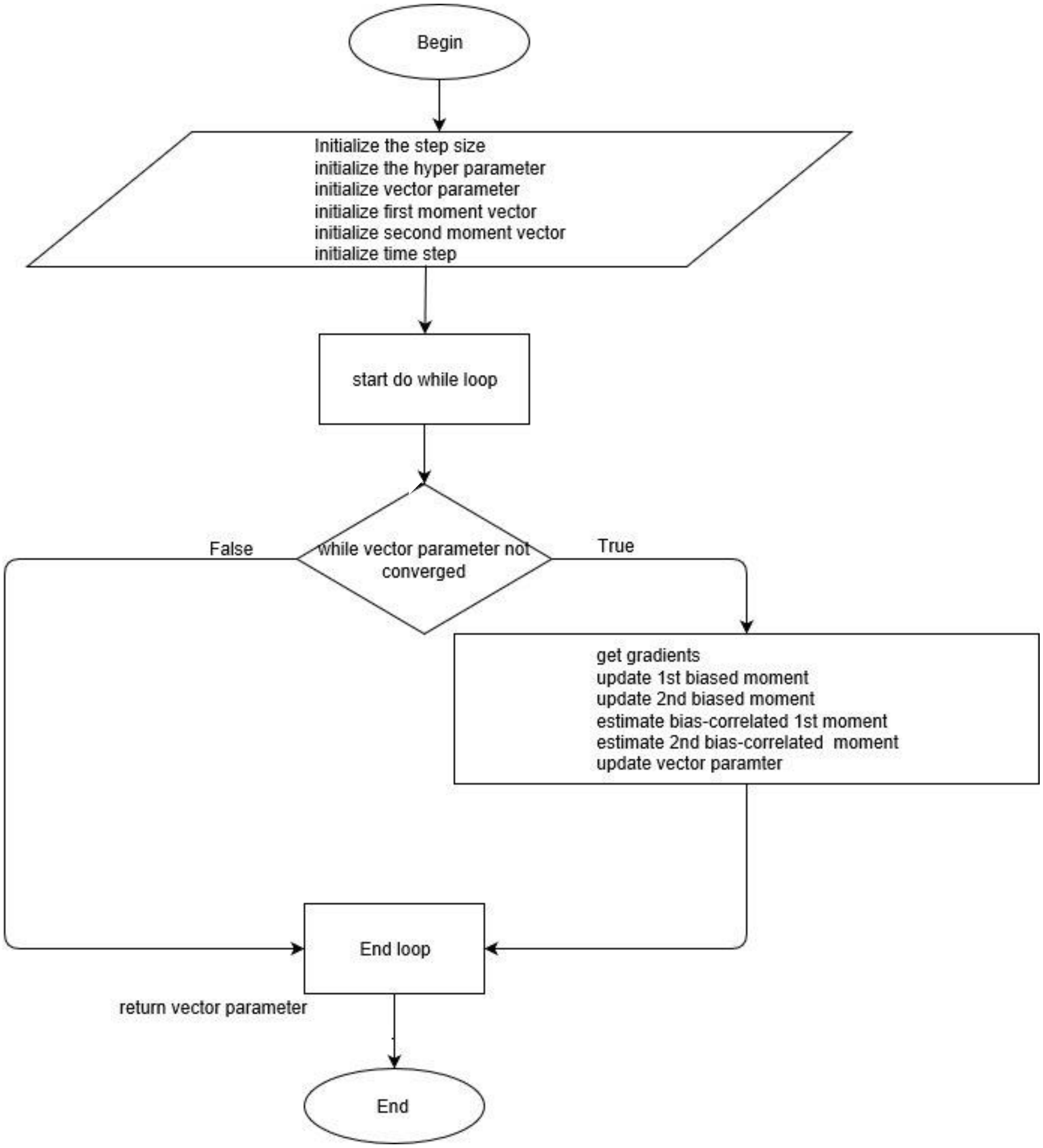

Figure 4: Flowchart of Proposed Algorithm

## 4. Results and Discussion

We have designed a convolutional neural network with autoencoder. The autoencoder is trained with 3 independent datasets with lung adenocarcinoma imaged with CT images. We have pre-processed the image through convolution filter and then passed the pre-processed image to the autoencoder for feature extraction. MIR is implemented to eliminate the reconstruction loss while reconstructing the input image, as a result increasing filter window size can decrease the processing time by allowing more data processing at once. The time taken during the process of cancer classification is measured based on the performance (classification). The processing time is measured with the unit frames per second. The processing time is calculated by adding the time taken to reconstruct the image and time taken to classify the image as shown in equation 9. The implementation was done by using Python version 2.7.0 using TensorFlow and model simulation was carried by taking 409 samples from 230 cancer patients, which were later divided into different transcriptome types. As the actual dataset image has very high resolution (20000 – 40000) pixels, we clipped the image into slices of dimension 2048*2048 for the input. The dataset that we collected has the frame rate averaging 10 frames/second. However, this frame rate is increased in our model by averaging 12 frames/second. We did not exclude any frames as our MIR model is able to process and filter the all-important features from that frame. The time taken during the process of cancer classification is measured based on the performance of the classification process. The processing time is measured with the unit frames per second. The processing time is calculated by adding the time taken to reconstruct the image and time taken to classify the image. The samples we have taken include different transcriptome types of cancer and the resolution for input image is taken 2048*2048. We implemented Python 3.7.0 version for extracting the image features. We have considered all subtypes transcriptome patterns such as micropapillary, papillary, acinar, and lepidic patterns. These patterns are tested in our work and the result is shown in the tables 2, 3, 4 and 5. For optimization, we have incorporated the Adam Algorithm. The MIR tool used Honglin Z, et al. [4] for image reconstruction can give better results, even if the filter window size is increased as shown in table 2.

During the feature extraction phase, the MIR tool can reconstruct the input image with all the important features, even if the input image quality is not as favorable as the RISA tool used in the state-of-the-art solution. Also, the results are compared according to filter window size. The MIR tool manages to increase the accuracy if the filter size increases, whereas the RISA tool decreases accuracy with a higher filter size.

| | | | | |
|---|---|---|---|---|
| Lepidic Pattern | 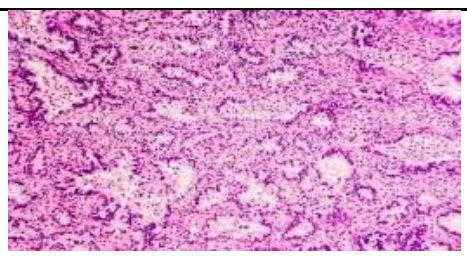 (a) | 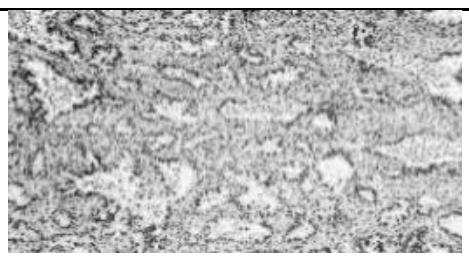 (b) | 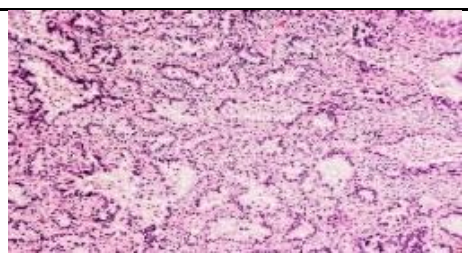 (c) | 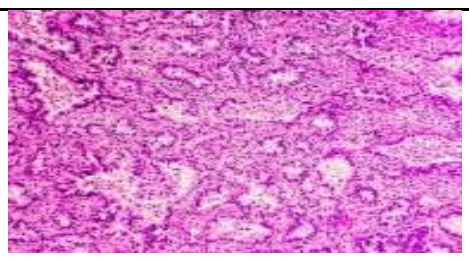 (d) |

Figure 5: Image Reconstruction sample (a) Original Image (b) Filtered image (c) State-of-the-art reconstructed image and (d) Proposed solution Reconstructed image

| | | | |
|---|---|---|---|
| 64*64 | 3*3 | 89.2 | 94.89 |
| | 4*4 | 62.5 | 70.56 |
| | 5*5 | 71.2 | 79.98 |
| 128*128 | 3*3 | 56.7 | 95.77 |
| | 4*4 | 35.9 | 78.56 |
| | 5*5 | 72.1 | 84.45 |

Figure 6: Performance based on filter size [1st column: image input size, 2nd Column: filter window size, 3rd column: State-of-art accuracy and 4th column: Proposed solution accuracy]

All the figures and tables of the proposed solution were compared to the state-of-the-art solution. Lung cancer image samples are analysed and presented in the Tables 2, 3, 4, and 5. The sample result shown in Figure 5 is achieved during the image reconstruction process. The result of table 2 is achieved by the varying the filter window size (effect of filter window size on accuracy). The results of Tables 3, 4, and 5 are obtained by varying the input image size. The result can provide accuracy and processing time during the extraction feature and classification stage. Reconstruction error is the inability of feature learning of the filter. It was considered as a parameter to calculate the accuracy whereas frames/second was considered the parameter for processing time. We have performed 39 comprehensives tests: 12 tests for each image input size of 128- and 512-pixel image and 18 tests for input size of 1028 and 2056-pixel image. Also, 9 tests were performed to test the filter window size (3*3, 4*4 and 5*5), as shown in Figure 6. Then, the result was calculated by taking the average of these results.

The computer used for data processing for this research is the Intel Pentium Silver N5000 Processor (1.1 GHz base frequency, up to 2.7 GHz burst frequency, 4 MB cache, 4 cores) with 8 GB DDR4-2400 SDRAM (1 x 8 GB) and 1 TB 5400 rpm SATA. The results for lung cancer classification were obtained and compared at different stages of image processing in the lung cancer classification system; reconstruction error is the main parameter of this result. By reducing the reconstruction error (by implementing the MIR tool), the accuracy of the proposed solution has improved lung cancer classification accuracy. By increasing the filter size, we were able to process a large amount of data in less time, which means increased frames per second. In Tables 2,3, and 4, the average classification accuracy is not quite improved because the input window size is low. However, by increasing the input window size, we were able to gain improved average classification accuracy. The confidence range provides the center of the range and confident values for our output metrics, that is, accuracy and processing time. The confidence interval is calculated by using Microsoft Excel and is measured in Python using TensorFlow with the confidence statistical function. Equation 8 calculates the confidence interval:

$$\mathrm{CI} = \bar{\mathrm{x}} \pm \mathrm{z}\frac{\sigma}{\sqrt{\mathrm{n}}} \tag{8}$$

Where:
CI is confidence range,
$\bar{x}$ is mean of the sample input,
z is fixed critical value for confidence integral,
σ is a standard deviation of the input data,
n is number of tests.

The formula for the standard deviation is:

$$\sigma = \sqrt{\frac{\sum|x-\bar{X}|^2}{n}} \tag{9}$$

σ = standard deviation
x = sample
$\bar{X}$ = mean of the sample
N = total number of samples

The proposed model has solved the limitations of the state-of-the-art model of this research. The proposed system achieves improved average classification accuracy of 99.5% against the state-of-the-art solutions, which has 98.9% classification accuracy. Also, our proposed model improves the average processing time to 12 frames per second against the state-of-the-art solution of 10 frames per second. See Figure 7 and Figure 8.

Table 2: Subtype Classification accuracy for state-of-art and proposed solution network and filter sizes,

| Window Size | Convolutional Window Size | RISA + Classifier Accuracy (%) (State-of-the-art) | MIR + Classifier Accuracy (%) (proposed Solution) |
|---|---|---|---|
| 32*32 | 3*3 | 52.5 | 59.67 |
| | 4*4 | 50.9 | 57.80 |
| | 5*5 | 71.3 | 76.88 |
| 64*64 | 3*3 | 89.2 | 94.89 |
| | 4*4 | 62.5 | 70.56 |
| | 5*5 | 71.2 | 79.98 |
| 128*128 | 3*3 | 56.7 | 95.77 |
| | 4*4 | 35.9 | 78.56 |
| | 5*5 | 72.1 | 84.45 |

Table 3: Accuracy and Processing time results of lung cancer Classification of dataset of Cancer Genome Atlas Database (CGAD)

| Dataset | Size of image (Pixels) | Input CT image | State-of-art (RISA and Classifier) | | | Proposed Solution (MIR reconstruction and Classifier) | | | Improved percentage of Accuracy |
|---|---|---|---|---|---|---|---|---|---|
| | | | | Accuracy % | Processing time (Frames/ Second) | | Accuracy % | Processing time (Frames/Second) | |
| Cancer Genome Atlas Database (CGAD) | 128 | 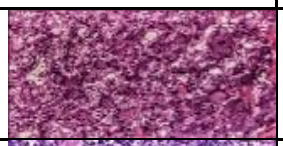 | 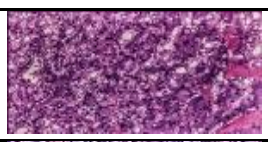 | 60.5 | 10 | 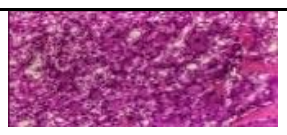 | 65.87 | 12 | 5.37 |
| | 512 | 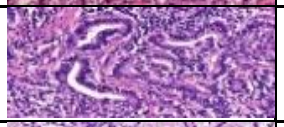 | 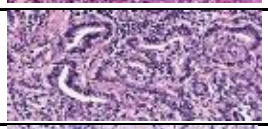 | 62.9 | 12 | 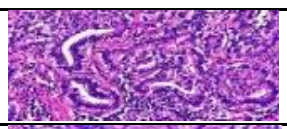 | 67.54 | 13 | 4.64 |
| | 1028 | 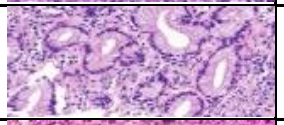 | 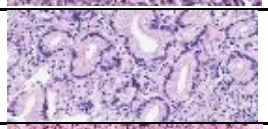 | 85. 5 | 11 | 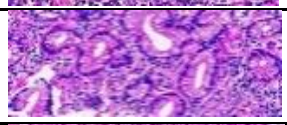 | 90.51 | 12 | 5.01 |
| | 2056 | 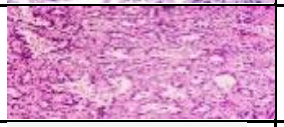 | 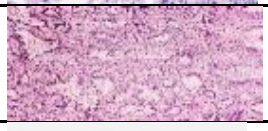 | 98. 9 | 12 | 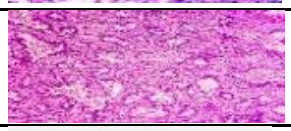 | 99.5 | 14 | 0.6 |
| | 128 | 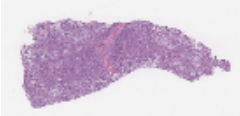 | 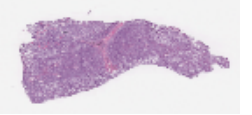 | 64.29 | 9 | 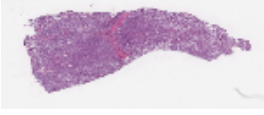 | 69.00 | 10 | 4.71 |
| | 512 | 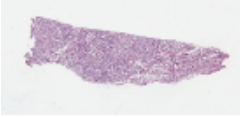 | 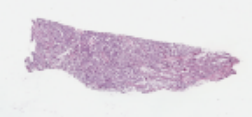 | 68.48 | 11 | 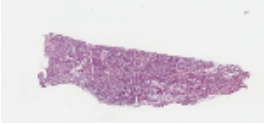 | 74.91 | 10 | 6.43 |
| | 1028 | 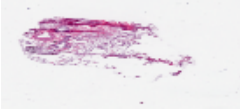 | 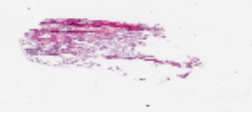 | 85.86 | 10 | 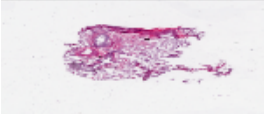 | 95.11 | 11 | 9.25 |
| | 2056 | 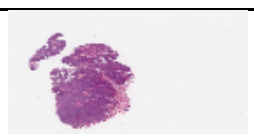 | 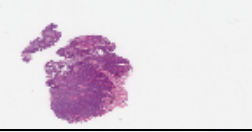 | 98.48 | 12 | 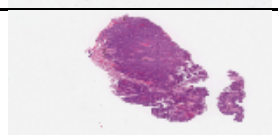 | 99.69 | 12 | 1.21 |
| | 1028 | 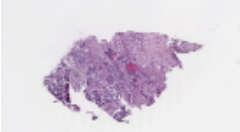 | 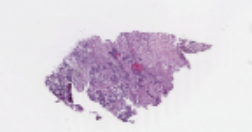 | 89.34 | 10 | 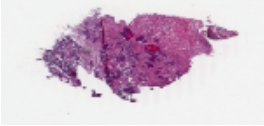 | 96.66 | 11 | 7.32 |
| | 2056 | 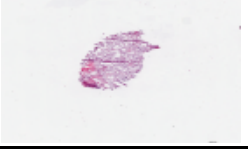 | 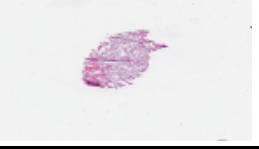 | 98.5 | 12 | 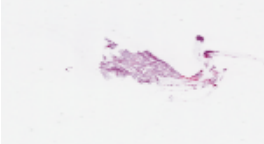 | 99.71 | 12 | 1.21 |

Table 4: Accuracy and Processing time results of lung cancer Classification of dataset of Lung CT segmentation Challenge (LCTSC)

| Dataset | Size of image (Pixels) | Input CT image | State-of-the-art (RISA and Classifier) | | | Proposed Solution (MIR reconstruction and Classifier) | | | Improved percentage of Accuracy |
|---|---|---|---|---|---|---|---|---|---|
| | | | | Accuracy % | Processing time (Frames/Second) | | Accuracy % | Processing time (Frames/Second) | |
| Lung CT segmentation Challenge (LCTSC) | 128 | 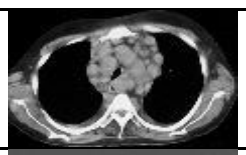 | 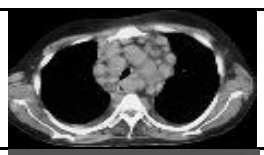 | 61.23 | 9 | 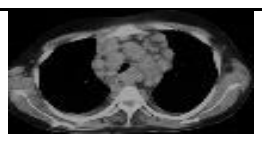 | 64.55 | 10 | 3.32 |
| | 512 | 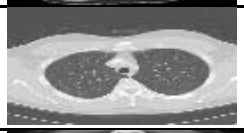 | 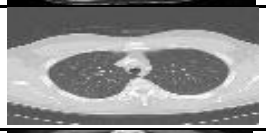 | 65.45 | 10 | 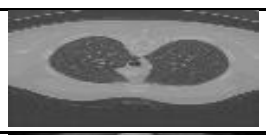 | 69.67 | 12 | 4.22 |
| | 1028 | 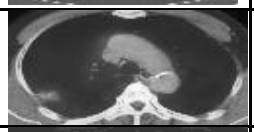 | 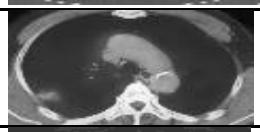 | 84. 5 | 12 | 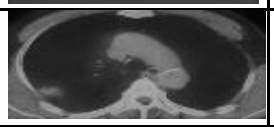 | 91.34 | 13 | 6.84 |
| | 2056 | 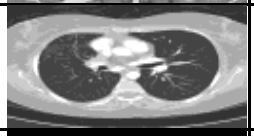 | 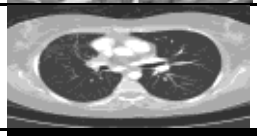 | 99.0 | 11 | 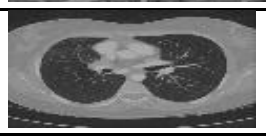 | 99.6 | 14 | 0.6 |
| | 128 | 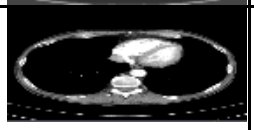 | 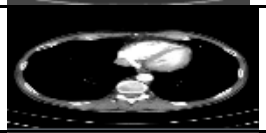 | 63.43 | 8 | 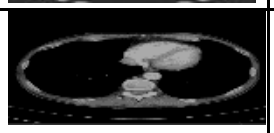 | 69.98 | 11 | 6.55 |
| | 512 | 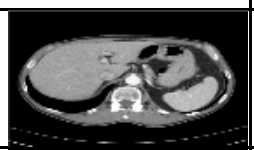 | 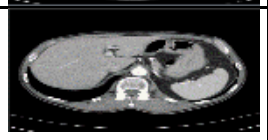 | 67.45 | 10 | 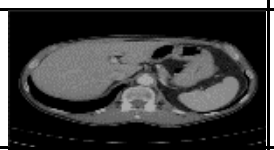 | 73.23 | 10 | 5.78 |
| | 1028 | 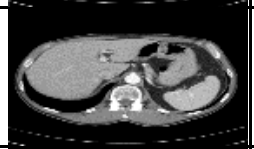 | 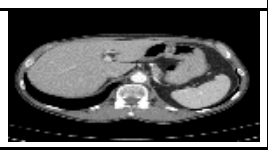 | 86.12 | 11 | 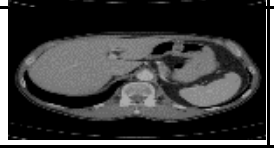 | 90.77 | 13 | 4.65 |
| | 2056 | 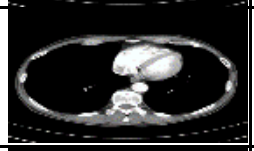 | 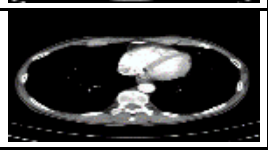 | 98.10 | 12 | 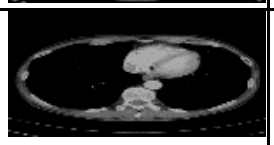 | 99.20 | 14 | 1.1 |
| | 1028 | 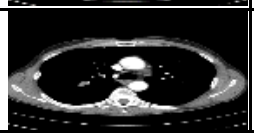 | 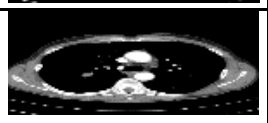 | 89.12 | 10 | 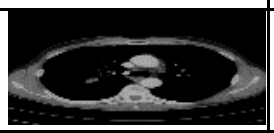 | 94.29 | 11 | 5.17 |
| | 2056 | 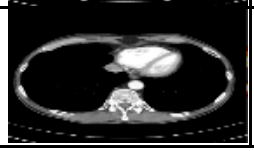 | 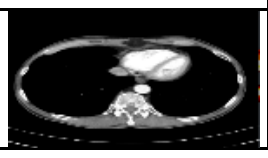 | 98.34 | 11 | 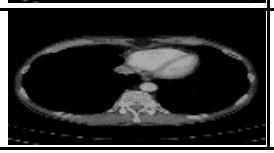 | 99.54 | 13 | 1.18 |

Table 5: Accuracy and Processing time results of lung cancer Classification of dataset of Lung CT segmentation Challenge (QLC)

| Dataset | Size of image (Pixels) | Input CT image | State-of-the-art (RISA and Classifier) | | | Proposed Solution (MIR reconstruction and Classifier) | | | Improved percentage of Accuracy |
|---|---|---|---|---|---|---|---|---|---|
| | | | | Accuracy % | Processing time (Frames/ Second) | | Accuracy % | Processing time (Frames/Second) | |
| QIN LUNG CANCER (QLC) | 128 | 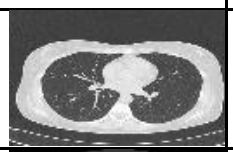 | 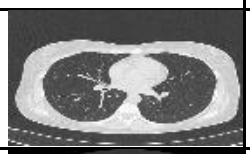 | 62.01 | 9 | 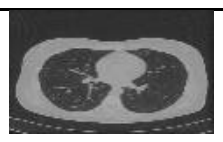 | 66.78 | 11 | 4.77 |
| | 512 | 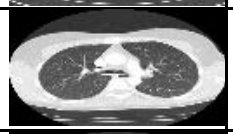 | 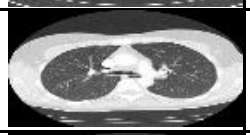 | 62.59 | 11 | 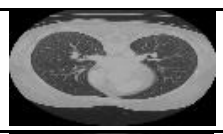 | 67.98 | 13 | 5.59 |
| | 1028 | 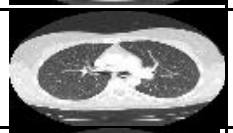 | 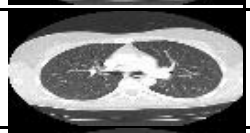 | 86.88 | 11 | 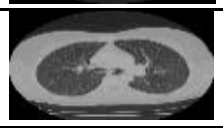 | 91.72 | 12 | 4.84 |
| | 2056 | 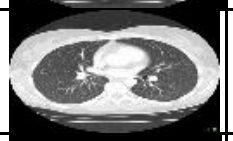 | 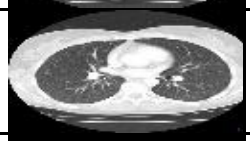 | 98. 7 | 12 | 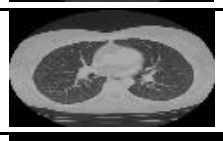 | 99.5 | 13 | 0.8 |
| | 128 | 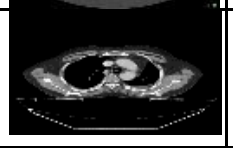 | 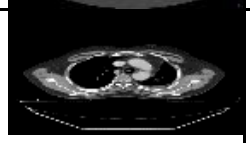 | 61.49 | 10 | 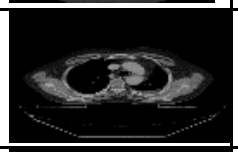 | 66.86 | 11 | 5.37 |
| | 512 | 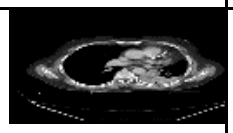 | 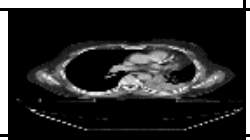 | 67.88 | 11 | 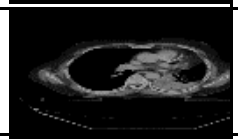 | 74.32 | 12 | 6.44 |
| | 1028 | 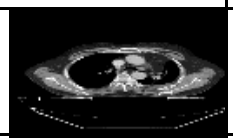 | 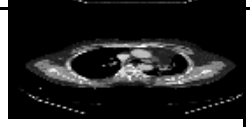 | 87.51 | 11 | 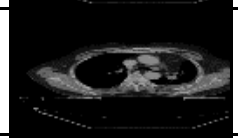 | 93.98 | 12 | 6.47 |
| | 2056 | 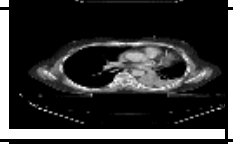 | 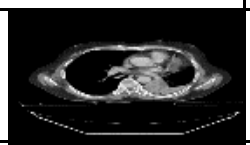 | 98.34 | 12 | 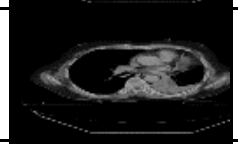 | 99.78 | 13 | 1.44 |
| | 1028 | 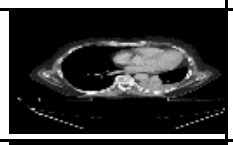 | 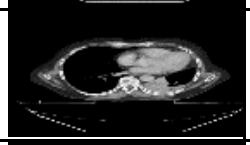 | 89.56 | 10 | 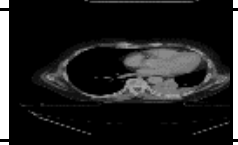 | 95.45 | 11 | 5.89 |
| | 2056 | 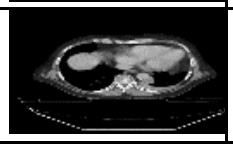 | 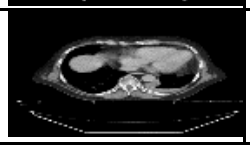 | 98.67 | 12 | 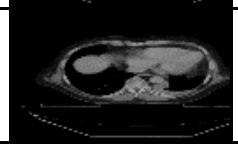 | 99.7 | 13 | 1.03 |

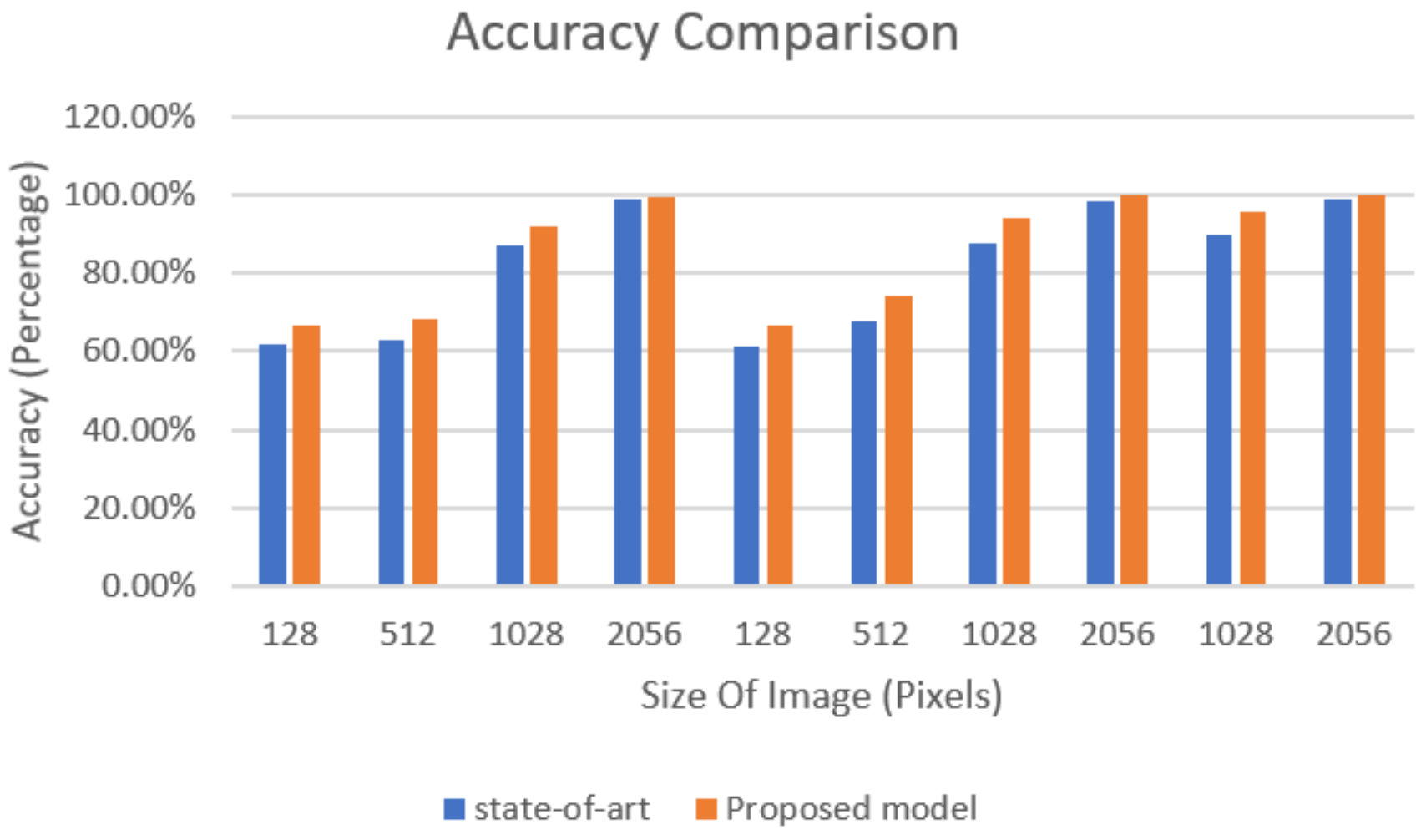

Figure 7: Classification Accuracy difference of state-of-the-art model and proposed solutions. This figure shows the average classification accuracy of state-of-the-art and proposed solutions. The first two bars show the classification accuracy of the image

input having size 128 pixel in percentage. The second two bars show the classification accuracy of the image input having size 512 pixel in percentage. The third two bars show the classification accuracy of the image input having size 1028 pixel in percentage. The fourth bar shows the classification accuracy of image input having size 2056 pixel in percentage and repeating the same procedure for more data.

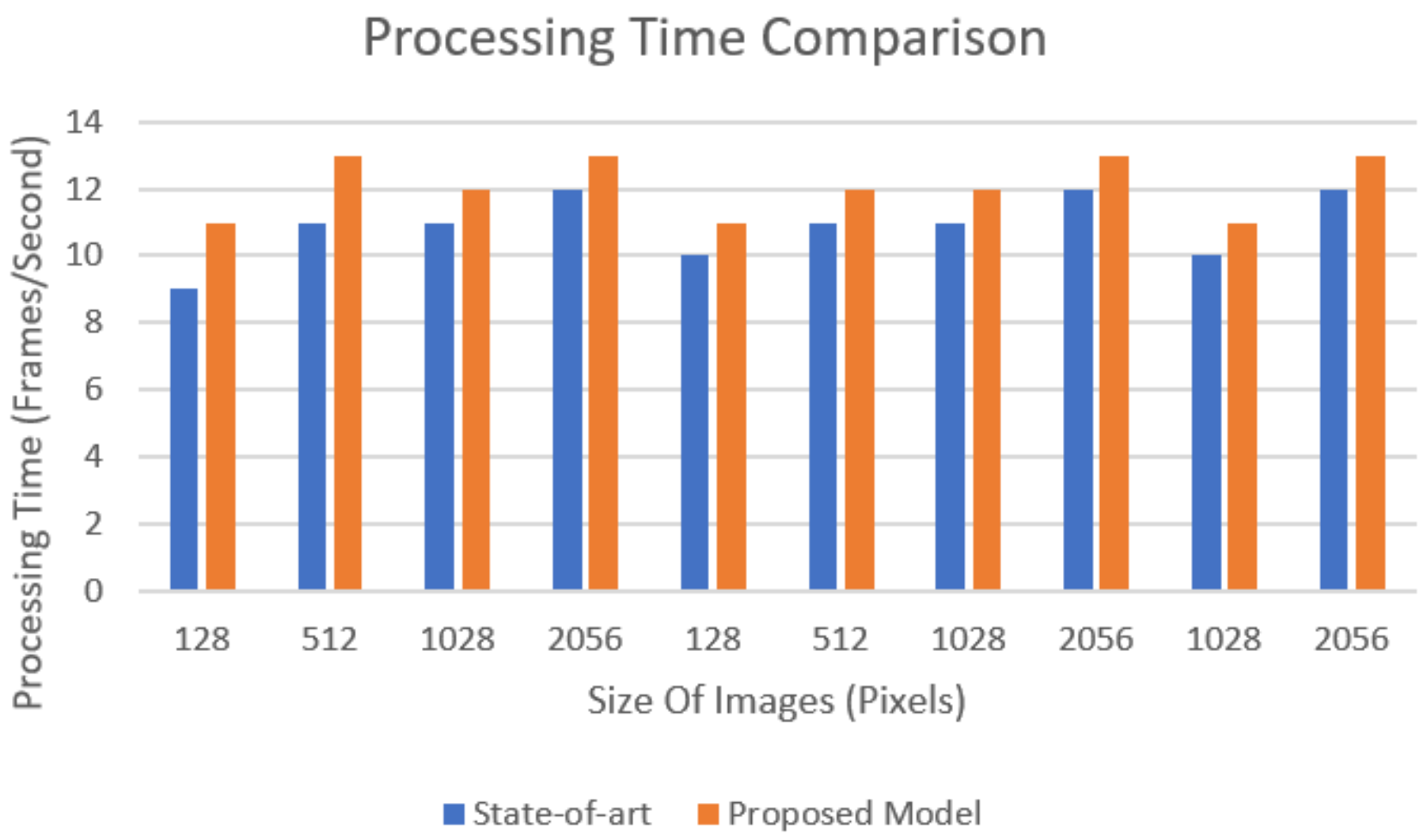

Figure 8: Processing Time difference of state-of-art and proposed solutions, it shows the average processing time of state of art and proposed solutions. Where, first two bar showing the processing time of image input having size 128 pixel in frames/second, second two bar showing the processing time of image input having size 512 pixel in frames/second, third two bar showing the processing time of image input having size 1028 pixel in frames/second and the fourth bar showing the processing time of image input having size 2056 pixel in frames/second and repeating the same process for more data.

Our results display the improvement in both classification accuracy and processing time comparing with the state-of-art solution, i.e. Antonio VAA, et al. [2]. The overall classification accuracy is increased by 0.6 % (98.9% to 99.5%) more than that of state-of-art solution with the help of MIR technique by allowing the use of any filter which reconstruct the image with minimum reconstruction error. Also, the processing time is decreased by 2 frames/second (10 frames/second to 12 frames/second) in our proposed system comparing to the state-of-art system with the help of MIR. Furthermore, less reconstruction error is obtained, and images are reconstructed in less time. The processing time of the proposed solution and state-of-art solution are obtained by running proposed solution code and state-of-art algorithm respectively. Multispace Image Reconstruction is the main feature of the proposed system used in feature extraction stage just after the image pre-processing stage which minimizes the reconstruction error which ultimately results in increased classification accuracy and decrease in processing time. Our proposed system uses python 3.7.0 version of programming language for implementing the proposed algorithm. The use of Multispace Image Reconstruction (MIR) does not only consider about RGB inputs, but it also composed of the gradient, GLCM, and LBP spaces. These results in minimum reconstruction error which then provides the high accuracy and less processing time. And use of Adam Optimization algorithm optimize the output minimizing processing time and increasing the classification accuracy. In conclusion, Convolutional Neural Network combined with autoencoder with MIR which enhanced the classification accuracy by 0.6 % more and reduce in processing time by 2 frames/second in overall. The limitations in state-of-art solution has been processed and refined by doing some research to solve the limitation. The proposed system solved the limitation with improved accuracy of 99.5% from the state-of-art accuracy 98.9%. The proposed system also reduces the processing time to 12 frames/second from 10 frames/second. These all improvements in accuracy and the processing time is due to MIR. The proposed system has better results in terms of accuracy and processing time in different sample image. Table 6 summarized the main features of this work, and gives a comparison between the proposed system and the state of art in term of the satisfied contributions and the proposed equations also.

Table 6: Comparison table of the State-of-the-art and Proposed Methods

| | **Proposed Solution** | **State-of-the-art Solution** |
|---|---|---|
| **Name of the solution** | Multispace Image Reconstruction (MIR) tool | Reconstructed image subspace analysis (RISA) tool |
| **Effect on accuracy** | Accurate and more clear reconstruction is performed through the MIR tool and it results in high classification accuracy of 99.5%. | Reconstructed Image is not clear, a blurry image is obtained and results in less classification accuracy. |
| **Effect on processing time** | Processing time is reduced from an average 10 frames/second to an average of 12 frames/second. | It provides an average processing time of 10 frames/second. |
| **Proposed equation** | $ML = MR + \lambda_s S$ <br> Where, <br> $MR = \sum_i^N \left(x_{i-1}^{out} - x_i^{in}\right)^2$- RL <br> $S = 1^n \sum_{j=1}^{M} \left(-r_j^{en} \log r_j^{en}\right)$ | $L = R + \lambda_s S$ <br> Where, <br> $R = \sum_i^N \left(x_i^{out} - x_i^{in}\right)^2$ <br> $S = 1^n \sum_{j=1}^{M} \left(-r_j^{en} \log r_j^{en}\right)$ |
| **Contribution 1** | Multispace Image Reconstruction (MIR) helps to reconstruct the image clearly. After reconstruction of the image, it is easy to classify the cancer as MIR provides ay clear reconstructed image. It reconstructs the image in different parts and integrates them for a better-reconstructed image. | Reconstructed image subspace analysis (RISA) technique cannot reconstruct the image clearly. The reconstructed image is blurred by RISA technique. |
| **Contribution 2** | Increasing the filter window size can decrease the processing time by allowing more data processing at once. | Was not used. |

# 5. Conclusion and Future Work

To classify the lung adenocarcinoma accurately, classification of CT images needs to be performed correctly. The proposed Deep learning for the classification of cancer has been successfully implemented. However, there are some limitations that affect the accuracy and processing time. The main purpose of this paper is to increase the overall prediction accuracy and reduce processing time by using MIR in the pooling layer of CNN by using the deep convolutional technique. MIR has been introduced in our research to reduce the reconstruction error, which is taken as a second-best solution [4]. MIR removes the blur by filtering the input CT image in a different space using different window size filters. It reconstructs the input images more clearly so that classification can be performed more accurately with a minimum reconstruction error. Therefore, the classification accuracy is improved by 0.6% and processing time is reduced by Antonio VAA, et al. [2] frames/second in average. For future research, a large dataset of CT images can be implemented and tested to extract the feature from a convolutional neural network and low pixel image input can be processed to evaluate the classification of cancer types. This method will be tested if it will enhance the performance and accuracy of the system or not.

## DECLERATION

No Funding for this work

No Conflicts of interests for this work

No Real Data has used